\documentclass[manuscript]{aastex}

\slugcomment{Not to appear in Nonlearned J., 45.}

\shorttitle{Twin Jets from the Z CMa Stellar System}
\shortauthors{Whelan et al.}

\begin{document}


\title{The 2008 Outburst in the Young Stellar System Z CMa: \\
The First Detection of Twin Jets}


\author{Whelan, E.T.\altaffilmark{1}}

\author{Dougados, C.\altaffilmark{1}}

\author{Perrin, M.D.\altaffilmark{2}}

\author{Bonnefoy, M.\altaffilmark{1}}

\author{Bains, I.\altaffilmark{3}} 

\author{Redman, M.P.\altaffilmark{4}}

\author{Ray, T.P.\altaffilmark{5}}

\author{Bouy, H.\altaffilmark{6}}

\author{Benisty, M. \altaffilmark{7}}

\author{Bouvier, J. \altaffilmark{1}}

\author{Chauvin, G. \altaffilmark{1}}

\author{Garcia, P.J.V. \altaffilmark{1,8}}

\author{Grankin, K. \altaffilmark{9}}

\author{Malbet, F. \altaffilmark{1}}


\altaffiltext{1}{Laboratoire dÕAstrophysique de Grenoble, UMR 5571, BP 53, 38041 Grenoble Cedex 09, France}
\altaffiltext{2}{Division of Astronomy, University of California, Los Angeles, CA 90095}
\altaffiltext{3}{Centre for Astrophysics and Supercomputing, Swinburne University of Technology, Hawthorn, Victoria 3122, Australia}
\altaffiltext{4}{Department of Physics, National University of Ireland Galway, Galway, Ireland.}
\altaffiltext{5}{School of Cosmic Physics, Dublin Institute for Advanced Studies, Dublin 2, Ireland}
\altaffiltext{6}{CAB (INTA-CSIC); LAEFF, PO Box 78, 28691 Villanueva de la Ca–ada, Madrid, Spain}

\altaffiltext{7}{INAF-Osservatorio Astrofisico di Arcetri, Largo E. Fermi 5, 50125 Firenze, Italy}
\altaffiltext{8}{Universidade do Porto, Faculdade de Engenharia, SIM Unidade FCT 4006, Rua Dr. Roberto Frias, s/n P-4200-465 Porto, Portugal}
\altaffiltext{9}{Crimean Astrophysical Observatory, 98409 Nauchny, Crimea, Ukraine}


\begin{abstract}
The Z CMa binary is understood to undergo both FU Orionis (FUOR) and EX Orionis (EXOR) type outbursts. While the SE component has been spectroscopically identified as an FUOR, the NW component, a Herbig Be star, is the source of the EXOR outbursts. The system has been identified as the source of a large outflow, however, previous studies have failed to identify the driver. Here we present adaptive optics (AO) assisted [FeII] spectro-images which reveal for the first time the presence of two jets. Observations made using OSIRIS at the Keck Observatory show the Herbig Be star to be the source of the parsec-scale outflow, which within 2\arcsec\ of the source shows signs of wiggling and the FUOR to be driving a  $\sim$ 0\farcs4  jet. The wiggling of the Herbig Be star's jet is evidence for an additional companion which could in fact be generating the EXOR outbursts, the last of which began in 2008 \citep{Grankin09}. Indeed the dynamical scale of the wiggling corresponds to a time-scale of 4-8 years which is in agreement with the time-scale of these outbursts. The spectro-images also show a bow-shock shaped feature and possible associated knots. The origin of this structure is as of yet unclear. Finally interesting low velocity structure is also observed. One possibility is that it originates in a wide-angle outflow launched from a circumbinary disk. 
\end{abstract}





\section{Introduction}

Z CMa is a unique pre-main sequence binary (d $\sim$ 1150 pc) whose behavior, over the last two decades, has captured the interest of astronomers studying outburst activity in young stars \citep{vanden04,szeifert10}. The binary is known to consist of a Herbig Be star (NW component) and an FU Orionis star (FUOR; SE component) with the binary position angle (PA) constrained at $\sim$ 130${^\circ}$ and separation at $\sim$ 0\farcs1 \citep{Barth94}. The classification of the SE component as an FUOR is based on its optical and infrared (IR) spectra which exhibit features associated with FUOR objects \citep{Hartmann89, Hartmann96}. For example in the optical the Balmer lines are found to have absorption features which are broad and blue-shifted by a few 100's of kms$^{-1}$. The NaI lines also show deep absorption features and in the IR strong CO absorption at 2.2$\mu$m is detected \citep{Bonnefoy10}. In addition to the recognised FUOR activity, Z CMa also undergoes EXOR-like outbursts, the source of which has recently been confirmed to be the Herbig Be star \citep{szeifert10}. EXOR are a class of eruptive variables, named after their class progenitor, EX Lupi and because they were originally believed to mimic FUORs \citep{Herbig89}. In the case of Z CMa, these $\sim$ 1$^{m}$-2$^{m}$ outbursts have typically occurred with a period of $\sim$ 5-10 years with recent events recorded in 1987, 2000, 2004 and 2008 \citep{vanden04, Grankin09}. 

Studies have also revealed rich outflow activity in the vicinity of Z CMa. \cite{Poetzel89} presented the first observations of the large-scale ($\sim$ 3.9 pc, PA =240$^{\circ}$) optical outflow. They also noted that the optical forbidden emission lines (FELs) tracing the jet close to the source had unusual multiple component profiles. In total they observe four distinct components within 12.5\arcsec\ of the central stars. A  component at $\sim$ -600 kms$^{-1}$ which extends to 2\arcsec\ , a component at -500 kms$^{-1}$ which extends to 12.5\arcsec\ , a component at -400 kms$^{-1}$ which extends to 6\arcsec\ and a final low velocity component ($<$ 100 kms$^{-1}$) which extends to 5\arcsec . Subsequent studies of the Z CMa outflow have spanned a large range in wavelength from the radio to the x-ray regimes and differed on whether the FUOR or the Herbig Be star is driving the outflow. For example \cite{Velazquez01} presented radio observations of a thermal jet component to the  Z CMa outflow. The jet had a PA of 245$^{\circ}$ in agreement with the \cite{Poetzel89} observations and the authors argue that the FUOR is the driver of the jet. \cite{Garcia99} published optical spectral images of the jet in the [OI]$\lambda$6300 line. They spatially associate the jet with the Herbig Be star rather than the FUOR however they also note the presence of the [OI]$\lambda$6300 line (an important jet tracer) in the spectrum of the FUOR. \cite{Stelzer09} observed with Chandra an x-ray jet with a PA of 225$^{\circ}$ $\pm$ 5$^{\circ}$.  Since the beginning of the last outburst (January 2008) we have been leading a large observational campaign aimed at better understanding Z CMa and in particular the origin of the EXOR outbursts \citep{Benisty10, Bonnefoy10}. 
An important aspect of this work are the observations focussed probing the outflow activity and its relationship if any to the outburst activity. In this letter we present remarkable new observations revealing the presence of two jets driven by the Z CMa  system. What is clear from these observations is that the Herbig Be star is the driver of the parsec-scale outflow (PA = 245$^{\circ}$) and that the FUOR is driving a 0\farcs4 micro-jet at a PA of $\sim$ 235$^{\circ}$.  

\section{Observations and Data Reduction}

The multiple component FEL profiles observed by \cite{Poetzel89} combined with the subsequent conflicting detections of the Z CMa jet strongly suggested to us the presence of more than one jet launched from within the  Z CMa system. With this in mind we used the OH-Suppressing Infra-Red Imaging Spectrograph ``OSIRIS" to investigate the two jets theory. OSIRIS is an integral field spectrometer (IFS) designed to work with the laser guide star and AO system on the W.M. Keck II telescope \citep{Wizinowich06}. It is a lenslet-based IFS which provides a spectral resolution of 3800 ($\sim$ 80 kms$^{-1}$) from 1-2.5 $\mu$m and offers a range of narrow and broad-band filters of varying pixel scales and hence fields of view (FOV) \citep{Larkin03}. The observations were taken on the night of December 22 2009 and in all cases the long-axis of the detector was set at a PA of 230$^{\circ}$. The AO system was used in NGS mode with Z CMa itself as the guide star. As we wished to focus on observing the Z CMa  [FeII] 1.257 $\mu$m and 1.644 $\mu$m jet emission we used the {\it Jbb} (1.18 to 1.416 $\mu$m) , {\it Hbb} (1.47 to 1.8 $\mu$m), {\it Jn2} (1.23 to 1.29 $\mu$m) and {\it Hn3} (1.59 to 1.68 $\mu$m)  filters.  The {\it Jbb} and  {\it Hbb} were chosen with the 20 mas pixel scale and the {\it Jn2} and the {\it Hn3} with the 35 mas scale.  The data was reduced using the OSIRIS data reduction pipeline which provides for the sky-subtraction, correction of various artifacts and the extraction of the individual spectra into a datacube. The cosmic-ray correction, continuum subtraction and mosaicing steps were done using specifically designed IDL routines. For further details on these procedures please refer to \cite{Amboage09}. Also note that the data has not been flux calibrated at this stage. As the {\it Hbb} and {\it Jn2} observations were found to be of lower S/N we present here the {\it Jbb} and {\it Hn3} results.  The FOV of the individual  {\it Jbb} and {\it Hn3} observations was 0\farcs32$\times$1\farcs28 and 1\farcs68$\times$2\farcs24 respectively. However as the source was positioned differently in each observation the mosaicing process increased the respective FOVs to approximately 0\farcs6$\times$1\farcs6 and 2\farcs4$\times$3\farcs4 . The seeing varied between 0\farcs75 and 0\farcs9 and the angular resolution achieved was 0\farcs05 for the {\it Jbb} observation and 0\farcs07 for the {\it Hn3}. Images obtained with the Suprime-Cam on the Subaru telescope and the SINFONI IFS are presented in Figures 2 and 3 respectively as a support to the OSIRIS results.





\section{Results and Discussion}

While the presence of two jets is immediately apparent from Figure 1 what is also evident is the rich and complex nature of the outflow activity in the vicinity of Z CMa. Hence we shall use the channel maps and position-velocity (PV) diagrams shown in Figures 1-4 to unravel and explore the origin of the different jet features. In all figures the contour levels start at 1$\%$-2$\%$ of the maximum and increase by factors of $\sqrt{2}$.  For the channel maps the velocity intervals are chosen to highlight the main features. The PV diagrams were generated with a pseudo-slit of 0\farcs1 ({\it Jbb}) to 0\farcs14 ({\it Hn3}) in width.  {\it All velocities are with respect to the stellar velocity taken at + 30 kms$^{-1}$(heliocentric)}. Below each set of structures are divided into Jet A, Jet B and a low velocity outflow and discussed separately.

 \subsection{Jet A: A Precessing Jet Driven by the Herbig Be Star}  
Jet A driven by the Herbig Be star is revealed in Figure 1 (b) at a PA of $\sim$ 245$^{\circ}$. The velocity and the PA of this jet approximately agree with what was imaged on the large-scale by \cite{Poetzel89} hence what is detected here is the sub-arcsecond initial section of the parsec-scale outflow. In Figure 2 (b) the jet is imaged on slightly larger scales and now extends to $\sim$ 1\arcsec\ with the addition of knot {\it K2}. Although {\it K2} is detected at a lower velocity we are confident that it is associated with the Herbig Be star jet as it lies along the same PA and is connected to the smaller scale jet  by extended emission along the jet channel.  From the PV diagram (Figure 4 a and b) the spread in jet velocities is $\sim$ 200 kms$^{-1}$. The most important feature of this jet is that it appears to be ``wiggling". This is especially clear in Figure 3 (a) where the image has been rotated so that the x-axis lies along the PA of Jet A. This behaviour extends to larger-scales as seen in Figure 3 (b). Figure 3 (b) is a section of a H$\alpha$ image taken with Suprime-Cam on the Subaru telescope.  The positions of knots 1-4 which correspond to the $\sim$ 30\arcsec\ jet first detected by \cite{Poetzel89} are slightly shifted to either side of the jet axis (by $\leq$ 1 \arcsec) offering further evidence of wiggling.

Jet precession is one possible explanation for jet wiggling \citep{Bally94, Chandler05}. One theory of jet precession considers a protostellar disk in a binary system where the disk is mis-aligned with the orbital plane of the binary. The tidal interaction between the disk and the companion causes the disk to precess which in turn causes precession in any jet launched from the inner regions of the disk \citep{Terquem99}. Alternatively the wiggling could result if the jet source is in orbital motion around a close companion \citep{Anglada07}. Either way the observed wiggling in Jet A offers strong evidence for an additional companion to the Herbig Be star. Accretion variability  in young stars while not precisely understood is generally attributed to an instability in the accretion disk which can be caused by the presence of a companion \citep{Forgan10}.  The typical size-scale of the observed wiggling is $\sim$ 0\farcs8 which for a jet with a velocity in the range 500-1000 kms$^{-1}$ corresponds to a time-scale or 4-8 years. This time-scale is in agreement with the time-scale of the EXOR outbursts hence a previously undetected companion to the Herbig Be star could explain the EXOR outbursts.

Finally we consider here the structures labelled as knot {\it K3} and a bow-shock shaped feature {\it B1} (Figure 2 a and b). The origin of {\it K3} is very uncertain. While it does align with Jet A and while the PV diagram shown in Figure 4 (b) shows emission linking knots {\it K3} to {\it K2} the close proximity of {\it K3} to {\it K2} is confusing and raises questions over whether it is in fact part of Jet A.  Additionally the source of {\it B1} is unclear. It is described here as being bow-shock shaped yet whether it is or not in fact a bow-shock needs to be confirmed through further observations. Also note that {\it K3} is coincident with the triangular-shaped {\it B1} the apex of which points into the Z CMa system. We offer two suggestions for the origin of these features. Firstly  {\it B1} could be a shock excited cavity where oblique shocks along the cavity walls combine to give its triangular shape with {\it K3} being part of the Herbig Be star's flow. Alternatively {\it B1} could in fact be a bow-shock which is part of a third jet driven by a nearby source.

\subsection{Jet B: A Collimated Micro-Jet Launched by the FUOR}
In Figure 1 (c) the [FeII] 1.257 $\mu$m emission associated with Jet B  is shown. The direction of the jet points to the position of the FUOR and therefore is clearly driven by this star. Also the jet is seen in the PV diagram Figure 4 (a) and has a velocity spread of $\sim$ 150 kms$^{-1}$. The PA of the FUOR jet agrees with the PA of the x-ray knot at 2 \arcsec\ \citep{Stelzer09} and with the bow-shock shaped feature labelled {\it D} located at 60 \arcsec\ \citep{Poetzel89} thus offering evidence of much older ejection events connected with the FUOR. Note that Jet B curves sharply at the end. This morphology is curious and it appears to be caused by a of mixing of the emission from Jet B  and knot {\it K1} (the origin of which is unclear).  Figure 1 (f) is a channel map extracted from SINFONI 1.257$\mu$m observations of Z CMa taken a year previous to the OSIRIS data \citep{Bonnefoy10}. While these observations were not optimised for the detection of jet emission the FUOR jet was observed. It is included here as the smaller pixel scale of the SINFONI observations (12.5 mas) means that the jet was traced closer to the FUOR position providing confirmation that it is indeed launched by the FUOR. 

Much evidence has been previously presented suggesting that FUORs drive outflows and collimated jets.  However the observations presented here are the first which unequivocally identify an FUOR as the driver of a collimated jet.   Difficulties with identifying FUOR outflows include uncertainties associated with the classification of the star as an FUOR and the identification of the driving source of the jet. As FUOR candidates are often embedded in reflection nebulae it can be difficult to say whether they or a perhaps unidentified close companion are the drivers of the observed jets. Our observations of Z CMa clearly illustrate that this is an important consideration. An example of another likely FUOR jets is the [SII] micro-jet associated with the probable FUOR PPS 13 \citep{Aspin01}. Detailed studies of FUOR jets will uniquely allow the impact of very high accretion rates on jet properties to be explored.  Current models predict scalings of jet parameters with mass accretion rate and whether or not these scalings hold as we move for example from low mass protostars to FUORs will potentially tell us something about the magnetisation of the FUOR disks \citep{Donati05}.

\subsection{Low Velocity Structure}
Distinct low velocity emission associated with the two jets is also detected. This is especially clear in the PV diagrams presented in Figure 4. In Figure 1 (b) and (c) we see that the emission associated with Jet A persists down to a velocity of $\sim$ 100 kms$^{-1}$ and is broader at lower velocities. Also Figure 1 (c) reveals the broader slower emission associated with the FUOR jet.  Classical T Tauri jets have been shown to have a nested structure with the fastest densest material located along the jet axis and surrounded by slower less collimated emission. This has been described as an ``onion-like" kinematic structure \citep{Bacciotti00, Lavalley00}. A proper analysis of the density, temperature and kinematics of these two jets is needed in order to confirm if this model can be applied here. In addition, an interesting low velocity feature is revealed in Figure 1 (d). A wide outflow with a velocity close to 0 kms$^{-1}$ and a PA intermediate between that of Jet A and Jet B is detected. Also it appears to originate from a position somewhere between that of the FUOR and the Herbig Be star. This structure presents itself in the PV diagrams as a distinct kinematical component and could be tracing a wide-angled flow launched from a circumbinary structure. Alternatively it could represent an expanding cavity pushed by the two jets. Further observations are needed to properly investigate the source of this component.

\section{Summary and Future Plans}
The observations of the Z CMa system presented here clearly highlight the complex nature of the outflow activity. In this letter we have begun the process of unravelling this activity. Firstly we have confirmed the existence of jets driven by both the FUOR and the Herbig Be star thus resolving the ambiguity over the driving source of the Z CMa parsec-scale outflow. Most interestingly we have observed the Herbig Be star jet to be wiggling. This strongly supports the existence of a second companion to the Herbig Be star. The idea that this companion could account for the EXOR outbursts must be investigated further. Finally we have demonstrated that FUOR can drive collimated jets and we are confident that future observations of this and other FUOR jets will provide important constraints to jet launching models. While these results provide new and significant information on the Z CMa system several questions remain, the most pressing being whether or not the implied third component to the system actually exists. In addition, further observational studies are planned in order to investigate the origin of the structures {\it K1}, {\it K3} and {\it B1}. It is envisioned that proper motion estimates will be particularly useful here. Considering the distance to Z CMa ($\sim$ 1150 pc) and the velocity of Jet A we would expect new features to emerge with a proper motion of $\sim$ 0\farcs1 per year. It would be also be interesting to observe over the next few years new ejection events which can be directly related to the last outburst. If emerging knots can be detected in Jet A this would absolutely tie this jet to the Herbig Be star and thus provide compelling evidence of the relationship between outflow and accretion activity. Finally a detailed study of the properties of Jets A and B (e.g. electron density, mass outflow rate) is currently underway and will be presented in a forthcoming paper. This will allow for comparison with leading jet launching models.

\acknowledgments{The authors would especially like to thank the OSIRIS instrument team and Keck Observatory staff for their guidance with the data acquisition and reduction. E.T. Whelan is supported by an IRCSET-Marie Curie International Mobility Fellowship in Science, Engineering and Technology within the 7th European Community Framework Programme. }



{\it Facilities:}  \facility{KECK (OSIRIS)}

\clearpage



 \begin{figure}
\includegraphics[scale=0.5]{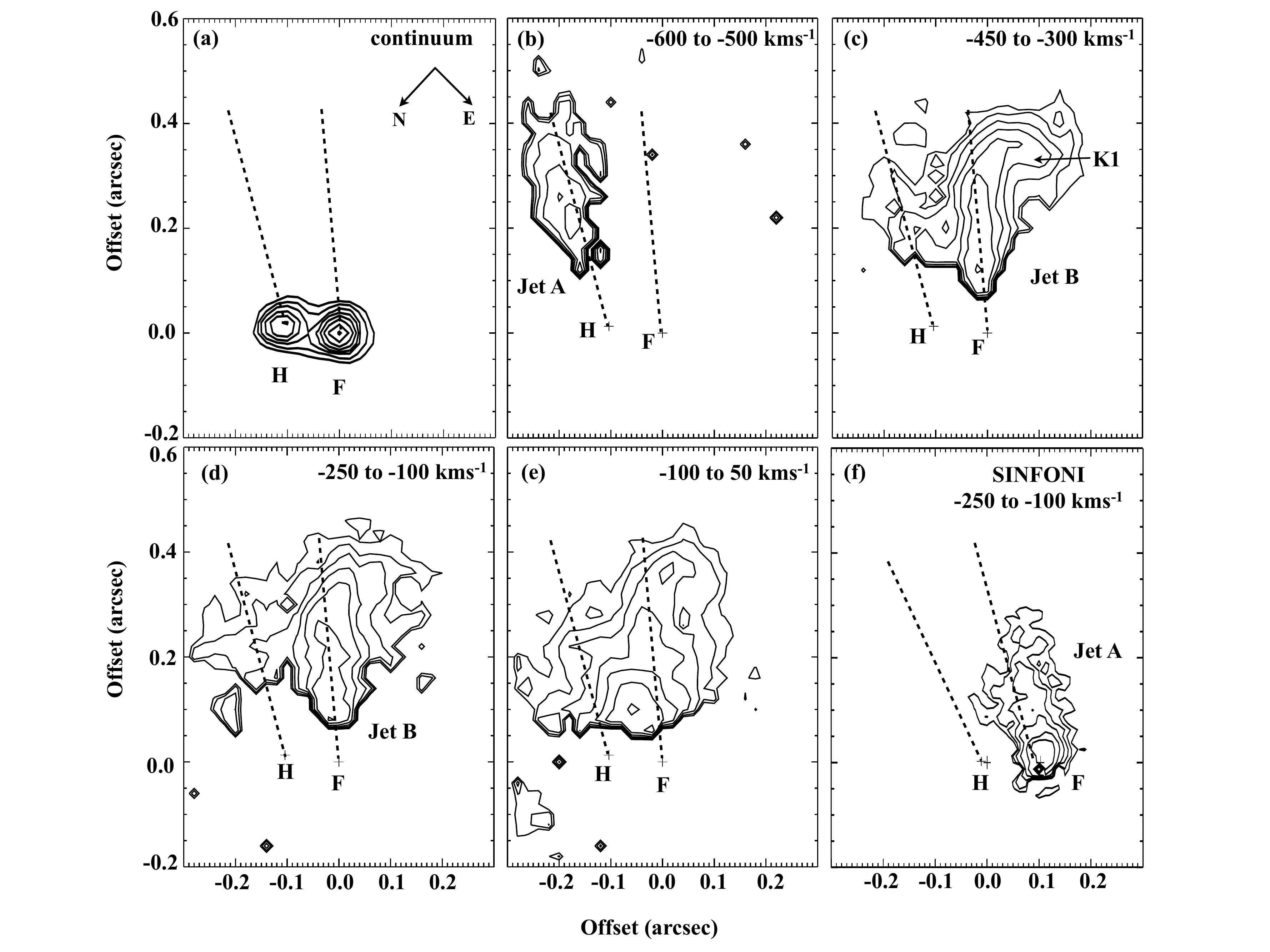}
 \caption{Continuum subtracted velocity channel maps of the Z CMa [FeII] 1.257 $\mu$m jet emission with the dashed lines delinating the PA of the jet structures. Contour levels start at 1$\%$-2$\%$ of the maximum and increase by factors of $\sqrt{2}$. The emission below S/N = 3 is masked. Jet A (PA $\sim$ 245$^{\circ}$) is revealed, the origin of which is the Herbig Be star (H). This jet shows signs of possible precession (also see Figures 2 and 3) which explains why the chosen PA does not correspond exactly to the jet axis (on the smallest scales). A lower velocity jet, Jet B at a PA of $\sim$ 235$^{\circ}$ is also seen. This jet extends to $\sim$ 0\farcs4 and is clearly driven by the FUOR (F).  Note the position of knot {\it K1} and the lower velocity emission associated with both jets. Panel (f) is a channel map extracted from a SINFONI observation obtained in December 2008. Due to the smaller pixel scale and shorter integration time only Jet B was detected. It is included here as it confirms that the FUOR is the driver of Jet B. {\it K1} is also detected in the SINFONI data. }
 \label{fone}
 \end{figure}

\clearpage


  \begin{figure}
\includegraphics[scale=0.5]{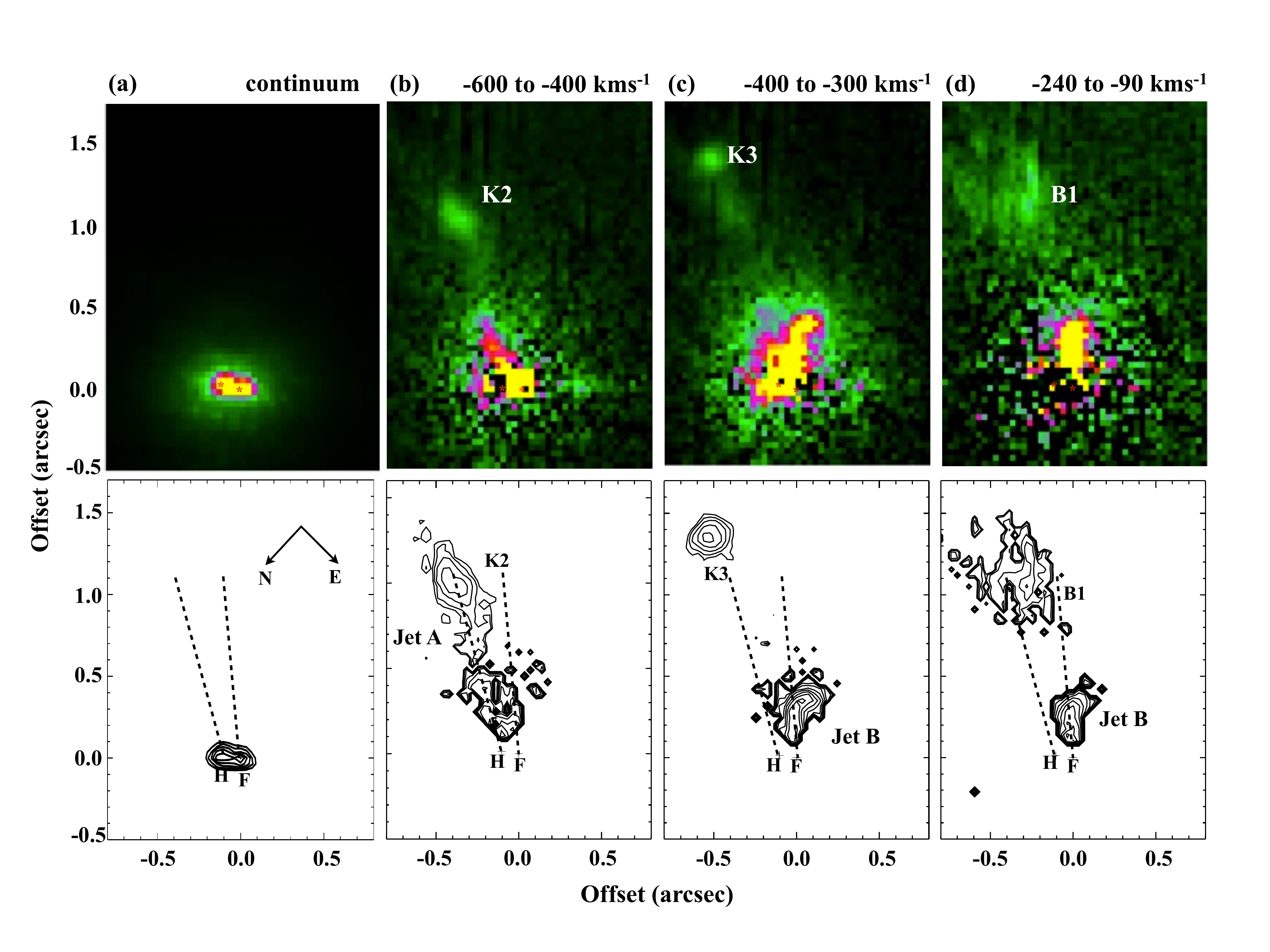}
 \caption{Continuum subtracted velocity channel maps of the Z CMa [FeII] 1.644 $\mu$m jet emission with the dashed lines delinating the PA of the jets. The extracted images are displayed on the top with the highest data values shown in yellow and the lowest in black. Jet A extends to $\sim$ 1 \arcsec\ with the addition of knot {\it K2}. The wiggling of  Jet A is very clear. A third knot {\it K3} and a bow-shock shaped feature {\it B1} are identified. Note the emission linking both {\it K2} and {\it K3} to the inner portion of Jet A (shown in Figure 1 b). To highlight the complicated structure (e.g. the wiggling in Jet A) the corresponding contour plots are displayed on the bottom. Again the contour levels start at 1$\%$-2$\%$ of the maximum and increase by factors of $\sqrt{2}$ and the emission below S/N = 3 is masked.}
 \end{figure}

\clearpage


 \begin{figure}
\includegraphics[scale=0.5]{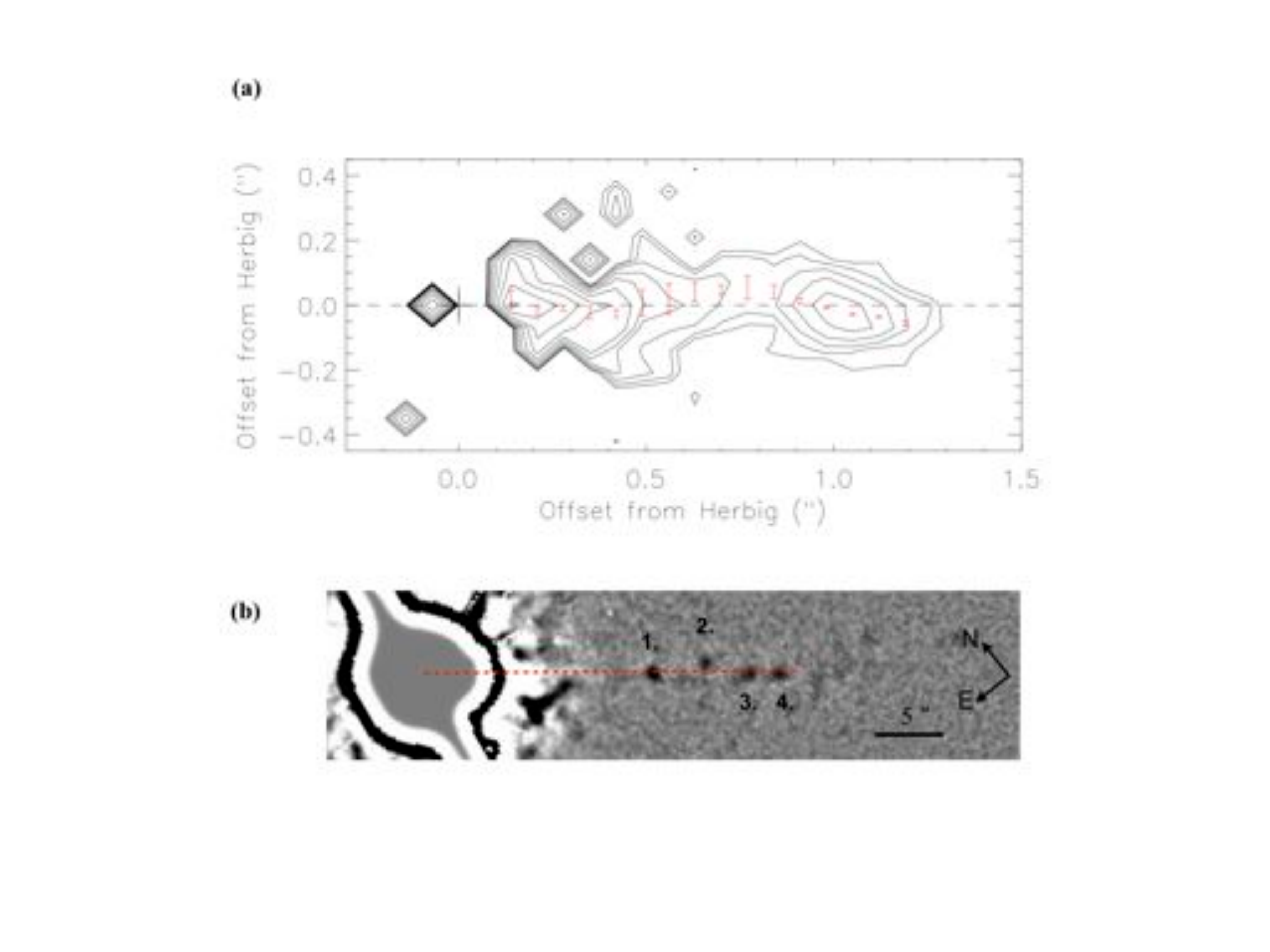}
 \caption{Figure illustrating the precession in Jet A. Top: a [FeII] 1.644$\mu$m channel map in the velocity range -600 kms$^{-1}$ to -500 kms$^{-1}$. The image has been rotated so that the x-axis corresponds to the jet axis. Contours start at 5$\%$ of the maximum and increase by $\sqrt{2}$. The wiggling of the jet about the jet axis is clear and we estimate the size-scale at $\sim$ 0\farcs8. Bottom: section from a H$\alpha$ image of the Z CMa region which shows the Jet A out to $\sim$ 30\arcsec. The pixel scale is 0\farcs2 and again the image is rotated so that the jet axis lies along the horizontal. The PA of these knots allows us to conclude that they are an extension of  Jet A. The image shows that the jet wiggling is seen further along the jet as the positions of the knots alternate around the jet axis by $\sim$ $\leq$ 1 \arcsec.}
 \end{figure}







\clearpage

\begin{figure}
\includegraphics[scale=0.5]{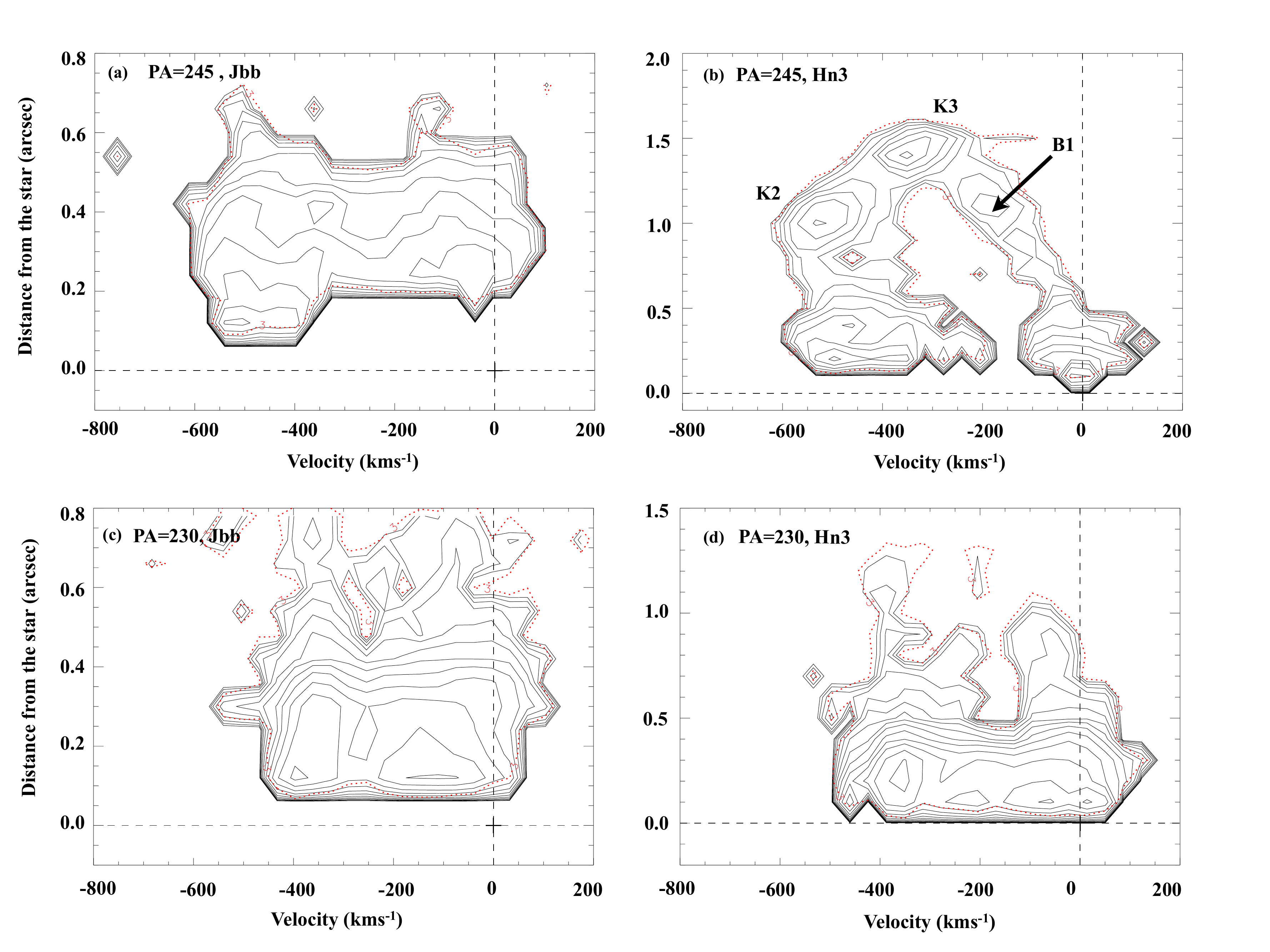}
 \caption{PV diagrams extracted from the {\it Jbb} and {\it Hn3} cubes at PAs corresponding to Jet A and Jet B. The contour levels start at 1-2 $\%$ of the peak and increase by $\sqrt{2}$. The emission below S/N =3 is masked with the dashed red line representing S/N = 3. The velocity components corresponding to the Herbig Be and FUOR jets are clearly visible. However what is especially noteworthy is the distinct lower velocity component associated with each jet. }
 \end{figure}




\end{document}